\documentclass[%
aps, prb,
reprint,
 amsmath,amssymb,
]{revtex4-1}

\usepackage{graphicx}% Include figure files
\usepackage{dcolumn}% Align table columns on decimal point
\usepackage{bm}% bold math
\usepackage{color}
\usepackage[utf8]{inputenc}
\usepackage[T1]{fontenc}
\usepackage{mathptmx}
\usepackage{upgreek}

\begin{document}

\preprint{AIP/123-QED}

\title[Integration and characterization of micron-sized YIG structures with very low Gilbert damping on arbitrary substrates]{Integration and characterization of micron-sized YIG structures with very low Gilbert damping on arbitrary substrates}

\author{P. Trempler}	
	\affiliation{Institut für Physik, Martin-Luther-Universität Halle-Wittenberg, D-06120 Halle, Germany}	
\author{R. Dreyer}%
	\affiliation{Institut für Physik, Martin-Luther-Universität Halle-Wittenberg, D-06120 Halle, Germany}
\author{P. Geyer}
	\affiliation{Institut für Physik, Martin-Luther-Universität Halle-Wittenberg, D-06120 Halle, Germany}
\author{C. Hauser}
	\affiliation{Institut für Physik, Martin-Luther-Universität Halle-Wittenberg, D-06120 Halle, Germany}
\author{G. Woltersdorf}
	\affiliation{Institut für Physik, Martin-Luther-Universität Halle-Wittenberg, D-06120 Halle, Germany}
\author{G. Schmidt}
	\email{georg.schmidt@physik.uni-halle.de}
	\affiliation{Institut für Physik, Martin-Luther-Universität Halle-Wittenberg, D-06120 Halle, Germany}
	\affiliation{Interdisziplinäres Zentrum für Materialwissenschaften, Martin-Luther-Universität Halle-Wittenberg, D-06120 Halle, Germany}

\date{\today}

\begin{abstract}
We present a novel process that allows the transfer of monocrystalline yttrium-iron-garnet microstructures onto virtually any kind of substrate. The process is based on a recently developed method that allows the fabrication of freestanding monocrystalline YIG bridges on gadolinium-gallium-garnet. Here the bridges' spans are detached from the substrate by a dry etching process and immersed in a watery solution. Using drop casting the immersed YIG platelets can be transferred onto the substrate of choice, where the structures finally can be reattached and thus be integrated into complex devices or experimental geometries. Using time resolved scanning Kerr microscopy and inductively measured ferromagnetic resonance we can demonstrate that the structures retain their excellent magnetic quality. At room temperature we find a ferromagnetic resonance linewidth of $\mu_0\Delta H_\mathrm{HWHM}\approx195\,\mathrm{\upmu T}$ and we were even able to inductively measure magnon spectra on a single micron-sized yttrium-iron-garnet platelet at a temperature of 5 K. The process is flexible in terms of substrate material and shape of the structure. In the future this approach will allow for new types of spin dynamics experiments up to now unthinkable.
\end{abstract}

\maketitle

\section{\label{sec:Introduction} Introduction}
The growth of high quality thin film yttrium-iron-garnet (YIG) is very challenging. Even today very low Gilbert damping ($\alpha\leq5\times10^{-4}$) is only achieved for deposition on gadolinium-gallium-garnet (GGG) which is almost perfectly lattice matched to YIG (see overview in Schmidt \textit{et.\,al.} \cite{Schmidt2020}). Nevertheless, for many experiments a GGG substrate is not suitable. GGG exhibits a strong paramagnetism that even \cite{Danilov1989} increases below $70\,\mathrm{K}$. This results in an enlarged Gilbert damping in a thin YIG film due to the coupling of the YIG with the substrate. As a consequence, many experiments which aim for example for the investigation of the strong coupling of magnons and microwave photons\cite{Huebl2013,Tabuchi2014} are limited to bulk YIG fabricated by liquid phase epitaxy (LPE) or to macroscopic YIG spheres. Up to now, this problem prevents experiments in hybrid quantum magnonics on YIG microstructures. Furthermore, experiments using YIG microstructures and integrated microwave antennae on GGG are difficult because of its large dielectric constant ($\epsilon\approx30$). Unfortunately, there also hasn't been any successful attempt to grow high quality YIG with reasonably low Gilbert damping on other substrates. Thus, a method to fabricate thin high quality YIG microstructures on GGG along a subsequent transfer on a different substrate would lead the way towards many new promising experiments and applications. We have developed a process that allows us to transfer YIG microstructures from GGG onto other substrates. Although the process is not suitable for mass fabrication it nonetheless enables a new class of experiments which until today seemed unthinkable.

\section{\label{sec:Processing} Processing}
Our method is based on a fabrication process \cite{Heyroth2019} using room-temperatue (RT) pulsed laser deposition (PLD), lift-off and annealing, which yields freely suspended YIG structures, whereby we apply the process in order to fabricate bridges or doubly clamped beams. The suspended parts of these structures exhibit extraordinary magnetic properties. For these structures a ferromagnetic resonance (FMR) linewidth at 9.6 GHz of $\mu_0\Delta H_\mathrm{HWHM}=140\,\mathrm{\upmu T}$ and a Gilbert damping of $\alpha\approx2\times10^{-4}$ were demonstrated.
\begin{figure}
\includegraphics[width=1\columnwidth]{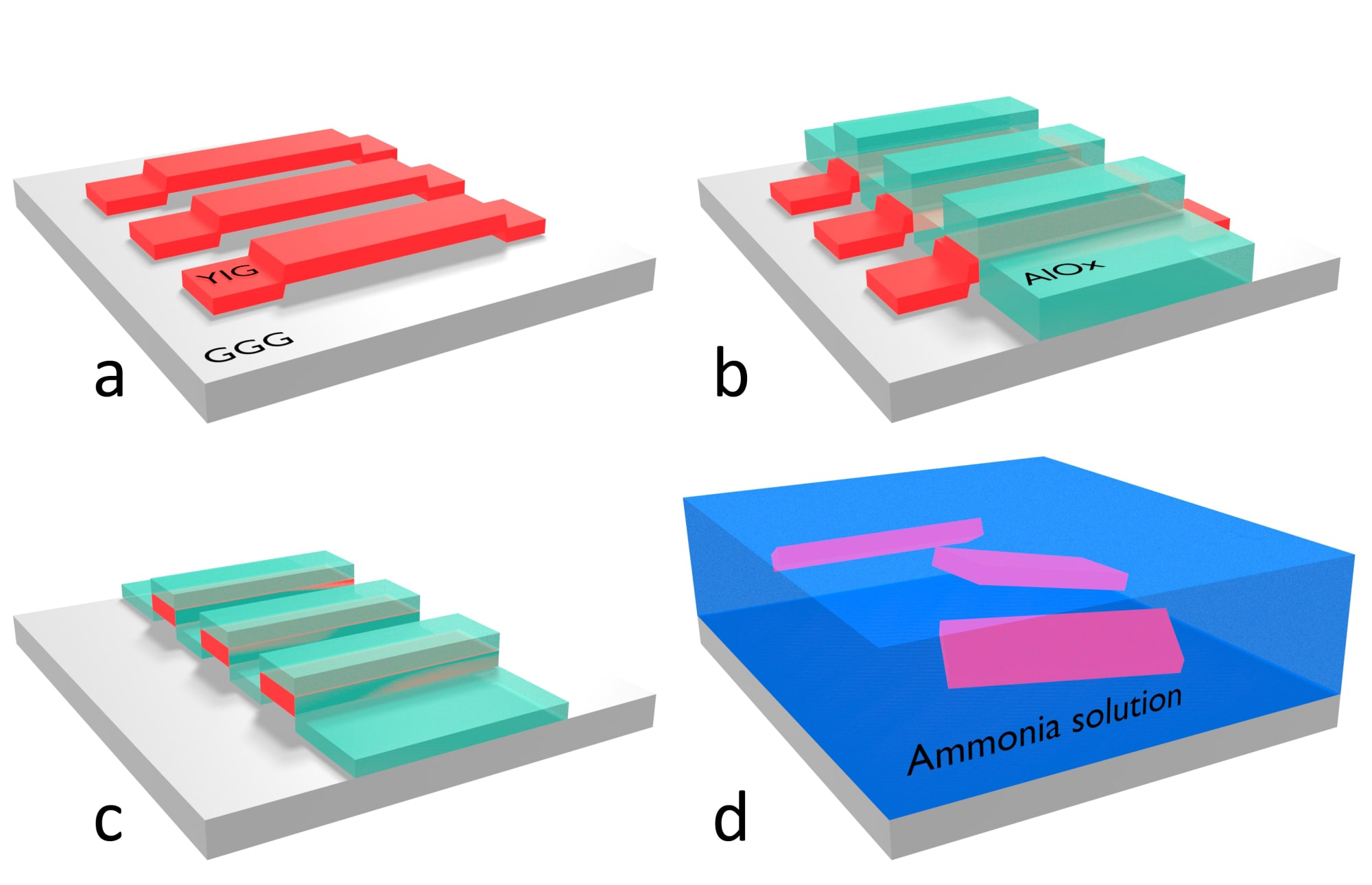}
\caption{\label{fig:transfer_wasser} Patterning process flow: (a) Array of monocrystalline YIG bridges \cite{Heyroth2019}. (b) The AlOx mask is deposited by e-beam lithography, evaporation, and lift-off. (c) The bridges are detached from the substrate by argon ion milling. (d) The AlOx is dissolved in ammonia water releasing the remaining YIG platelets into the liquid.}
\end{figure}
Using this process we fabricate an array of 500,000 bridges of $1.5 \times 5\,\mathrm{\upmu m^2}$ span-size on a GGG substrate. We then mask the spans of the bridges by aluminum oxide using electron beam lithography, e-beam evaporation, and lift-off. [Fig.\,\ref{fig:transfer_wasser}\,(b)]. Using argon ion milling allows to remove the part of the bridge that connects the span to the substrate leaving the masked YIG as a micro slab like platelet embedded in the aluminum oxide (AlOx). [Fig.\,\ref{fig:transfer_wasser}\,(c)]. Dissolving the mask in ammonia water lifts the 500,000 YIG micro platelets from the substrate and immerses them in the solution. The wet etchant is then stepwise replaced by water yielding a watery suspension of uniform monocrystalline YIG platelets [Fig.\,\ref{fig:transfer_wasser}\,(d)]. By drop-casting the YIG platelets can now be transferred to any substrate. After drying, the platelets stick to the substrate and even stay in place during subsequent spin-coating of further resist layers. With the help of additioanl lithography the platelets can be integrated in complex devices or applications.

\begin{figure}
\includegraphics[width=0.8\columnwidth]{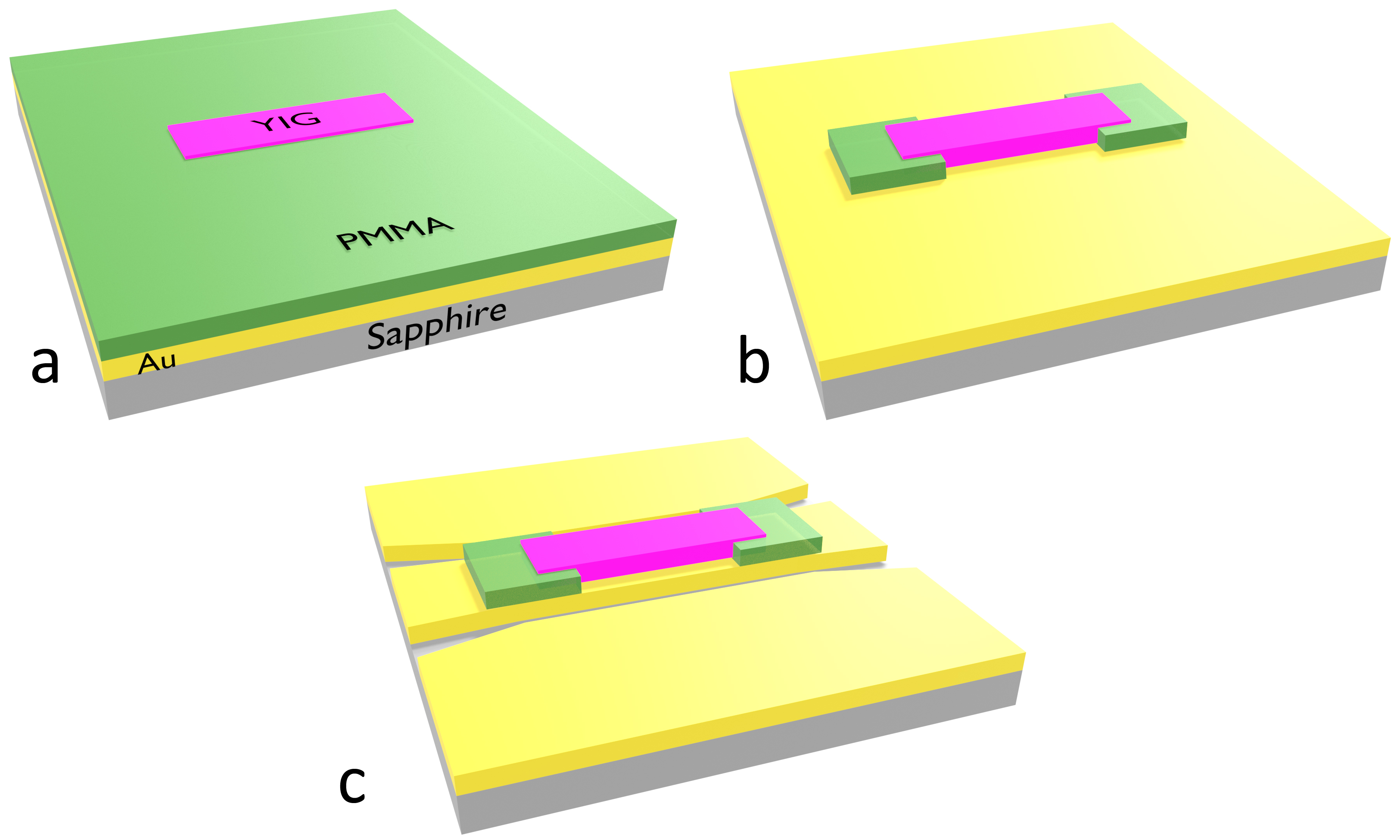}
\caption{\label{fig:cpw} (a) The YIG drop-cast on the PMMA sinks into the polymer during heating. (b) The PMMA at the ends of the platelet is crosslinked to fix the YIG to the Au. (c) Electron beam lithography and dry etching are used to pattern the CPW.}
\end{figure}

\begin{figure}
\includegraphics[width=1\columnwidth]{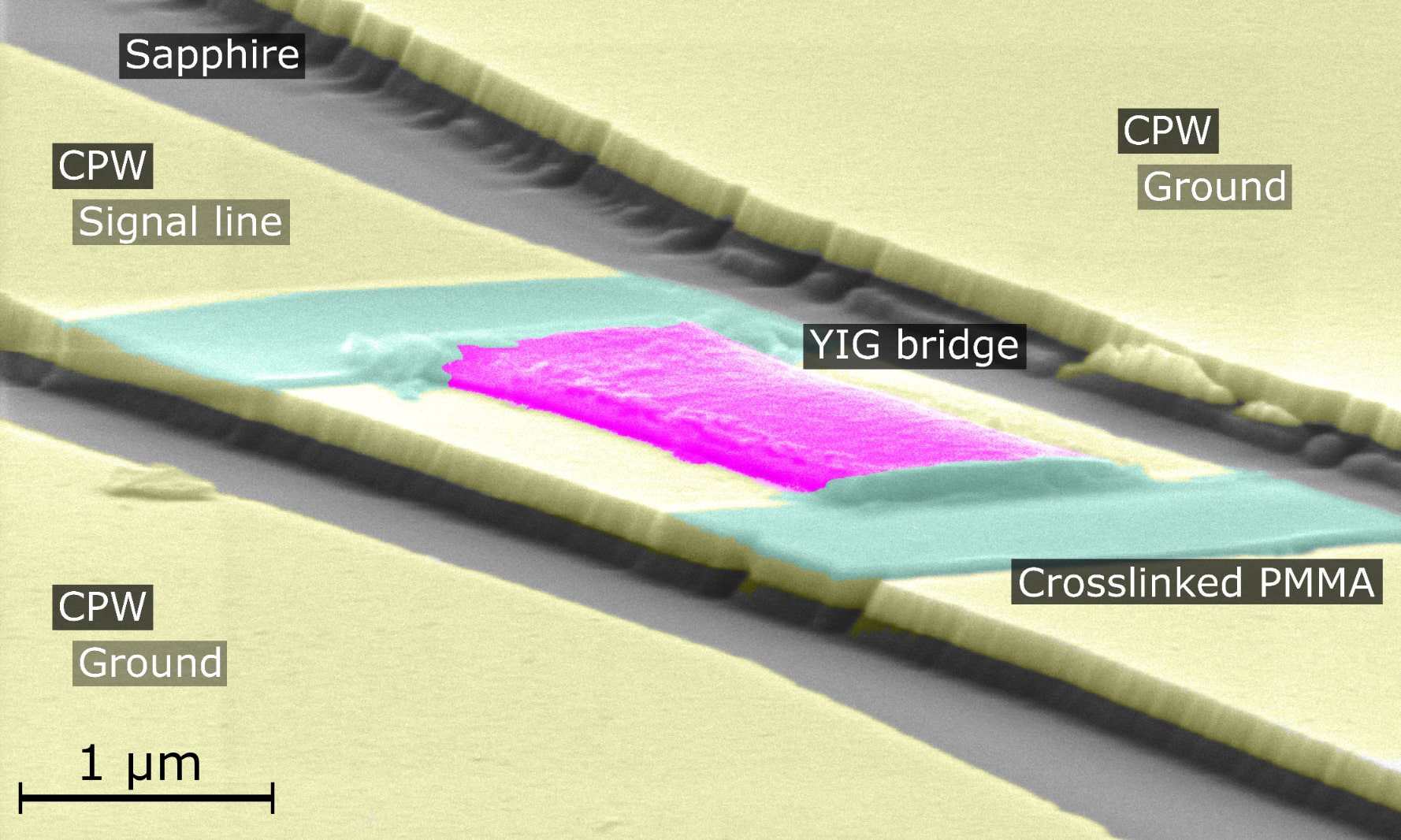}
\caption{\label{fig:bridge_on_CPW} False-color SEM image of a transferred YIG platelet (magenta) fixed with crosslinked PMMA (green) on top of a Ti/Au CPW (yellow). The bridge has a span length of $4.5\,\mathrm{\upmu m}$, a width of $1.5\,\mathrm{\upmu m}$ and a nominell YIG layer thickness of approximately $160\,\mathrm{nm}$.}
\end{figure}

Here we show one example how a YIG platelet can be integrated into a coplanar waveguide geometry to achieve in-plane excitation and high sensitivity in FMR. As a substrate we use sapphire onto which $150\,\mathrm{nm}$ of Au with a Ti adhesion layer were deposited by electron beam evaporation. Sapphire is chosen because of its excellent properties for high frequency measurements. Before the drop-casting, a layer of PMMA is spun onto the sample. The suspension is exposed for a few seconds to ultrasonic agitation to ensure a homogeneous suspension of the YIG platelets and by using a pipette a single drop of the suspension is then put onto the sample. After the drop-casting the YIG platelets are typically flat on the sample surface but randomly oriented. Once a suitable YIG platelet is identified we heat the sample up to $250\,^\circ \mathrm{C}$ which is well above the glass transition temperature of the PMMA \cite{Mohammadi2017} causing the YIG platelet to slightly sink into the PMMA film [Fig.\,\ref{fig:cpw}\,(a)]. By electron beam lithography we then crosslink the PMMA at the end of the bridge, defacto welding the bridge to the Au surface [Fig.\,\ref{fig:cpw}\,(b)]. Using the PMMA layer under the YIG has several advantages compared to direct deposition on the Au surface. No spin coating is required before the bridge is fixed and after removing the non-crosslinked PMMA the sample surface is now also clean from possible residue of the drop-casting process. It should be noted that there is most likely a gap of $10-40\,\mathrm{nm}$ between YIG and Au so the system corresponds rather to a bridge with a YIG platelet as a span and two pedestals of PMMA as posts. To realize the final test structure we now use electron beam lithography, AlOx evaporation and lift-off to mask the intended area of the CPW and the YIG platelet itself. By Argon ion milling we remove the unmasked Au and Ti. After removing the AlOx mask we end up with a CPW perfectly aligned with the YIG platelet and ideally suited in terms of size and shape for the FMR characterization of the YIG platelet [Fig.\,\ref{fig:cpw}\,(c)]. The final structure is shown in Fig.\,\ref{fig:bridge_on_CPW} as an false-color SEM image.

\section{Magnetic properties}
In order to assess the sensitivity of our experiment we now perform FMR measurements. The samples are bonded onto a sample holder that fits into a $^4$He bath cryostate. The cryostate is placed inside an electromagnet that can be rotated in the sample plane. The external magnetic field can be modulated using an air coil of a few turns of Cu wire wound around the sample holder inside the cryostate. For our measurements the external magnetic field is oriented along the long side of the platelet. RF excitation is done by applying an RF signal with a power of $-21\,\mathrm{dBm}$. Measurements are performed by sweeping the magnetic field at constant RF frequency. The transmitted RF signal is rectified and the modulation of the external field allows for lock-in detection to increase sensitivity. With the YIG platelet centered on the waveguide the exciting RF field is oriented in the sample plane and homogeneous over the YIG platelet. As a consequence we can only excite standing spin wave modes with an uneven number of antinodes that have non-zero magnetization.

\begin{figure}
\includegraphics[width=1\columnwidth]{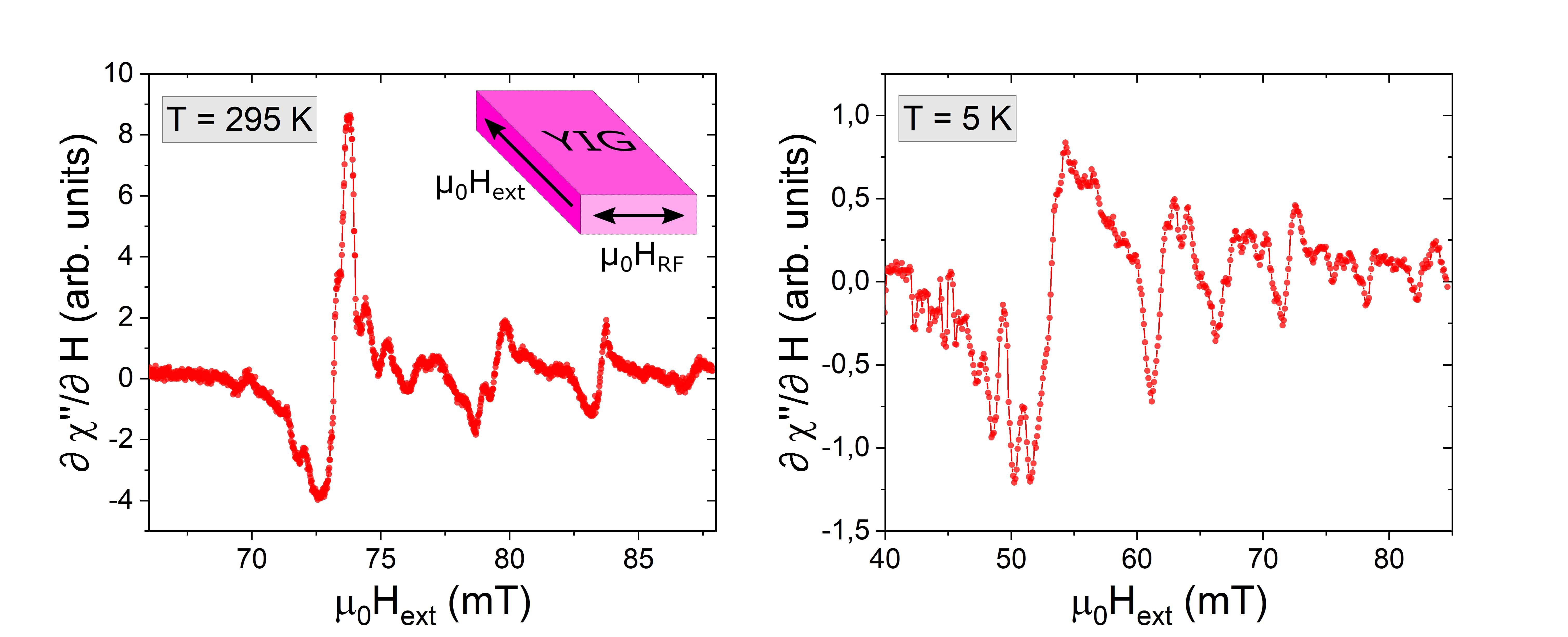}
\caption{\label{fig:FMR_Messung_single_Bridge} FMR spectra for a frequency of $4\,\mathrm{GHz}$ at (a) $5\,\mathrm{K}$ and (b) $295\,\mathrm{K}$ showing the occurance of several spin wave modes in the YIG bridge. The extended spin-wave spectra even for low temperatures suggests a very low Gilbert damping.}
\end{figure}

\begin{figure}
\includegraphics[width=1\columnwidth]{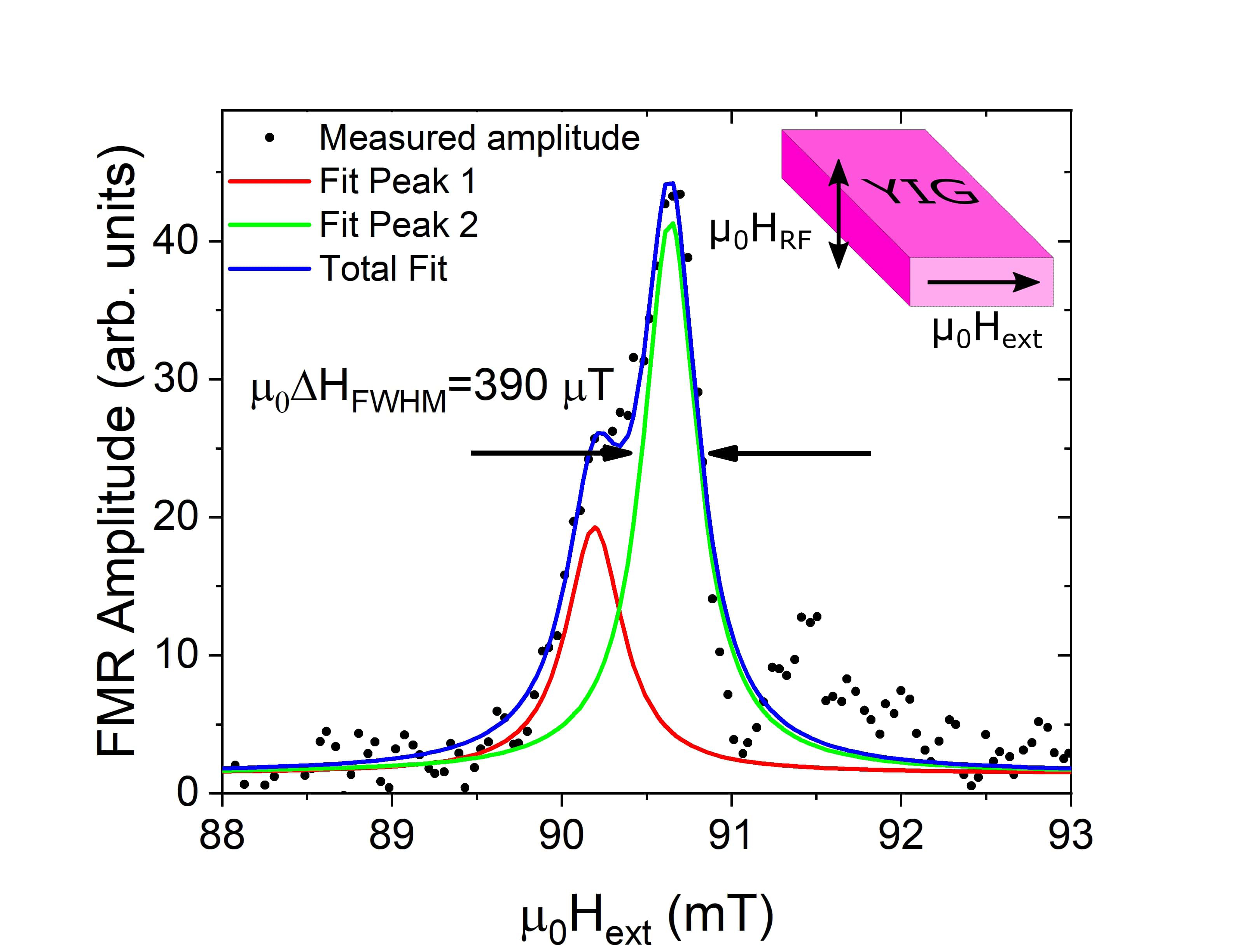}
\caption{\label{fig:line} Main FMR line as composition of two separate lines for a single transferred YIG platelet of $1.5\times4.5\,\mathrm{\upmu m^2}$. The linewidth is $\mu_0\Delta H_\mathrm{HWHM}=195\,\mathrm{\upmu T}$.}
\end{figure}

Fig.\,\ref{fig:FMR_Messung_single_Bridge} shows two resonance curves obtained at $4\,\mathrm{GHz}$ at room temperature and at $5\,\mathrm{K}$ respectively. In both cases we observe an extended spin-wave spectrum with a large number of backward-volume modes (BVMs). These discrete modes are caused by the finite size of the YIG platelet and correspond to standing spin wave modes as observed in a previous experiment \cite{Heyroth2019}. Because of the complexity of the spectrum and the overlap of multiple modes it is difficult to obtain a linewidth or even extract a Gilbert damping from measurements at different respective frequencies. A closer look at the shape of the main resonance line indicates that it is not a single line but composed from at least two separate lines if not more [Fig \ref{fig:line}]. At $5\,\mathrm{K}$ the spectrum is more noisy than at room temperature but still the details of the spectrum are similar to those at room temperature. The major difference to the room temperature measurement is the change in resonance field that can be attributed to the change in saturation magnetization \cite{HansenRoeschmannTolksdorf1974}.

\begin{figure}
\includegraphics[width=1\columnwidth]{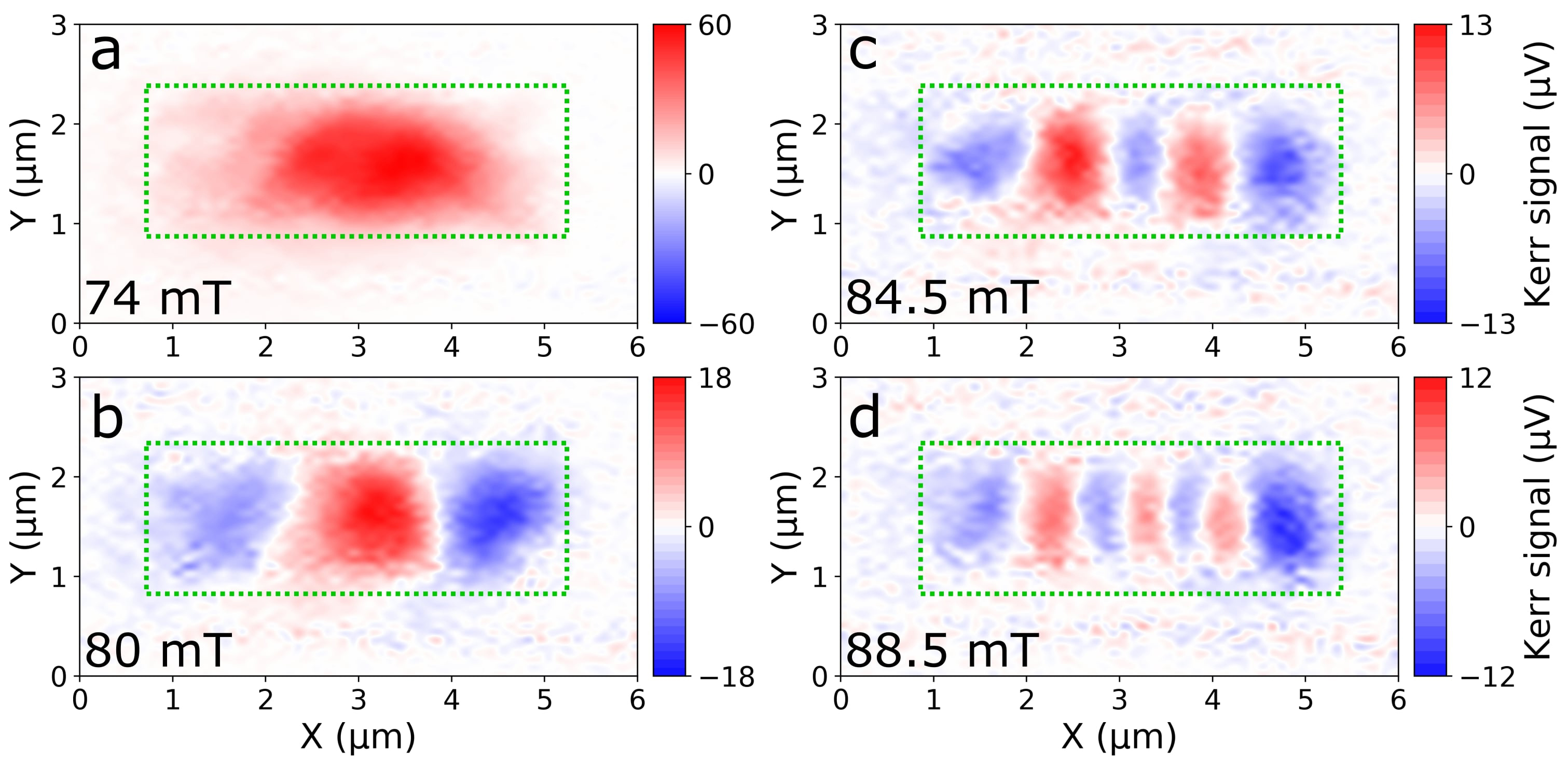}
\caption{\label{fig:MOKE_Messung_single_bridge} Spatial resolved measurements acquired at a frequency of $4\,\mathrm{GHz}$ at different respective magnetic fields. The TRMOKE images show standing BVMs in the span of the bridge for $\mu_0 H_\mathrm{ext}$ of (a) $74\,\mathrm{mT}$, (b) $80\,\mathrm{mT}$, (c) $84.5\,\mathrm{mT}$ and (d) $88.5\,\mathrm{mT}$. The dotted lines serve as a guide to the eye to indicate the approximate sample position. $\mu_0 H_\mathrm{ext}$ is applied along the x-direction.}
\end{figure}

We perform TRMOKE experiments on the YIG in order to obtain more detailed information about the local structure of the excited modes. Further details of this technique are described in the work of Tamaru \textit{et.\,al.}\cite{Tamaru2002} and Neudecker \textit{et.\,al.} \cite{Neudecker2006}. Again the measurements are performed with the external magnetic field oriented along the long side of the platelet. TRMOKE allows to locally image magnon modes in terms of both intensity and phase \cite{Heyroth2019}. To perform the spatially resolved imaging the frequency was set to 4 GHz at an RF amplitude of -25 dBm. The real and imaginary part of the dynamic susceptibility were detected in pointwise fashion while the magnetic field was kept constant for each picture [Fig \ref{fig:MOKE_Messung_single_bridge}].

The spatially resolved measurements show several standing BVM with the fundamental mode with only one antinode [Fig.\,\ref{fig:MOKE_Messung_single_bridge}\,(a)] and three standing BVMs with antinodes distributed along the bridge in Fig.\,\ref{fig:MOKE_Messung_single_bridge}\,(b)-(d) \cite{Heyroth2019}. As expected all observed modes exhibit an uneven number of antinodes. Again, it is not possible to extract a precise value for the line width for this sample. Another platelet from the same batch was transferred into the gap of a coplanar waveguide. In this geometry the out-of-plane RF field allows for TRMOKE measurements with the external field applied perpendicular to the long side of the platelet. This results in a larger spacing between the resonance lines and yields the spectrum shown in Fig.\,\ref{fig:line}. The resonance field is slightly shifted compared to the measurements shown in Fig.\,\ref{fig:MOKE_Messung_single_bridge}. At 4 GHz we observe two superimposed lines which can be fitted by two lorentzian line shapes. We obtain a linewidth of $\mu_0\Delta H_\mathrm{HWHM}\approx 195\,\mathrm{\upmu T}$. To the best of our knowledge even for large area thin films there are only two publications from other groups that show a smaller linewidth at this frequency \cite{Kelly2013, Ding2020}. For untransferred bridges (on GGG) we have already measured a smaller linewidth, however, it is unclear whether the original sample produced for the drop-casting was of similar quality. In any case the magnetic quality is only weakly affected by the transfer, if at all.

\section{Outlook}
The presented process opens up a large number of options. As we have shown in \cite{Heyroth2019} the 3D patterning process is not limited to linear bridges. Besides we can also make frames, rings, circular drums, tables, or other arbitrariliy shaped flat structures which would allow us to use the transfer technique presented here. The main restriction is merely the size. With increasing structure size the yield of the initial 3D patterning process is reduced and also the writing time increases linearly with the area. On the other hand we need a large number of structures to have enough statistical hits in the drop-casting process. A low concentration of YIG structures in the suspension would make the drop-casting a hopeless procedure. Beyond that, are even more options. Before the masking with AlOx we can perform additional processing on the bridges. We can for example deposit a thin metal film on top. After detachment and drop-casting we have a 50:50 chance that the metal film ends up at the bottom of our platelet. A second evaporation step could then be used to create a double side metallized YIG film as has been used in \cite{Li2016} for the demonstration of magnon drag. In our case, however, we have no limitations as to the metals that we want to use and their respective thicknesses. Furthermore we may even be able to nanopattern the metal before detaching the bridges and finally achieve a piece of YIG thin film with lithographically nanopatterned metal on both sides. Our structures may even be suitable for hybrid quantum magnonics at $\mathrm{mK}$ temperatures. As van Loo \textit{et.\,al.} \cite{Loo2018} and Mihalceanu \textit{et.\,al.} \cite{Mihalceanu2018} have shown, the damping of thin film YIG increases at low temperatures, mainly because of interaction with the GGG substrate. In our case the YIG platelet is no longer on the substrate. Even more it has never been in direct contact with GGG so also contamination effects can be excluded, making high performance at $\mathrm{mK}$ temperatures even more likely. And finally these isolated structures may also be suitable for the formation of magnon-based Bose-Einstein condensates\cite{Demokritov2006}.

\section{Conclusion}
We have demonstrated that it is possible to transfer high quality thin film YIG microstructures onto other substrates and to integrate them in complex experiments. The magnetic quality is only slightly affected by the process, if at all. \\ Notably, we are able to measure FMR spectra at $5\,\mathrm{K}$ with many details. This process opens up new routes towards a multitude of experiments which formerly seemed completely out of reach.

\section{Data availability}
The data that support the findings of this study are available from the corresponding author upon reasonable request.

\begin{acknowledgments}
We wish to acknowledge the support of TRR227 project B02 WP3 and project B01.
\end{acknowledgments}


\begin{thebibliography}{15}%
\makeatletter
\providecommand \@ifxundefined [1]{%
 \@ifx{#1\undefined}
}%
\providecommand \@ifnum [1]{%
 \ifnum #1\expandafter \@firstoftwo
 \else \expandafter \@secondoftwo
 \fi
}%
\providecommand \@ifx [1]{%
 \ifx #1\expandafter \@firstoftwo
 \else \expandafter \@secondoftwo
 \fi
}%
\providecommand \natexlab [1]{#1}%
\providecommand \enquote  [1]{``#1''}%
\providecommand \bibnamefont  [1]{#1}%
\providecommand \bibfnamefont [1]{#1}%
\providecommand \citenamefont [1]{#1}%
\providecommand \href@noop [0]{\@secondoftwo}%
\providecommand \href [0]{\begingroup \@sanitize@url \@href}%
\providecommand \@href[1]{\@@startlink{#1}\@@href}%
\providecommand \@@href[1]{\endgroup#1\@@endlink}%
\providecommand \@sanitize@url [0]{\catcode `\\12\catcode `\$12\catcode
  `\&12\catcode `\#12\catcode `\^12\catcode `\_12\catcode `\%12\relax}%
\providecommand \@@startlink[1]{}%
\providecommand \@@endlink[0]{}%
\providecommand \url  [0]{\begingroup\@sanitize@url \@url }%
\providecommand \@url [1]{\endgroup\@href {#1}{\urlprefix }}%
\providecommand \urlprefix  [0]{URL }%
\providecommand \Eprint [0]{\href }%
\providecommand \doibase [0]{http://dx.doi.org/}%
\providecommand \selectlanguage [0]{\@gobble}%
\providecommand \bibinfo  [0]{\@secondoftwo}%
\providecommand \bibfield  [0]{\@secondoftwo}%
\providecommand \translation [1]{[#1]}%
\providecommand \BibitemOpen [0]{}%
\providecommand \bibitemStop [0]{}%
\providecommand \bibitemNoStop [0]{.\EOS\space}%
\providecommand \EOS [0]{\spacefactor3000\relax}%
\providecommand \BibitemShut  [1]{\csname bibitem#1\endcsname}%
\let\auto@bib@innerbib\@empty
%</preamble>
\bibitem [{\citenamefont {Schmidt}\ \emph {et~al.}(2020)\citenamefont
  {Schmidt}, \citenamefont {Hauser}, \citenamefont {Trempler}, \citenamefont
  {Paleschke},\ and\ \citenamefont {Papaioannou}}]{Schmidt2020}%
  \BibitemOpen
  \bibfield  {author} {\bibinfo {author} {\bibfnamefont {G.}~\bibnamefont
  {Schmidt}}, \bibinfo {author} {\bibfnamefont {C.}~\bibnamefont {Hauser}},
  \bibinfo {author} {\bibfnamefont {P.}~\bibnamefont {Trempler}}, \bibinfo
  {author} {\bibfnamefont {M.}~\bibnamefont {Paleschke}}, \ and\ \bibinfo
  {author} {\bibfnamefont {E.~T.}\ \bibnamefont {Papaioannou}},\ }\href
  {\doibase 10.1002/pssb.201900644} {\bibfield  {journal} {\bibinfo  {journal}
  {physica status solidi (b)}\ }\textbf {\bibinfo {volume} {257}},\ \bibinfo
  {pages} {1900644} (\bibinfo {year} {2020})}\BibitemShut {NoStop}%
\bibitem [{\citenamefont {Danilov}\ \emph {et~al.}(1989)\citenamefont
  {Danilov}, \citenamefont {Lyubon'ko}, \citenamefont {Nechiporuk},
  \citenamefont {Ryabchenko} \emph {et~al.}}]{Danilov1989}%
  \BibitemOpen
  \bibfield  {author} {\bibinfo {author} {\bibfnamefont {V.}~\bibnamefont
  {Danilov}}, \bibinfo {author} {\bibfnamefont {Y.~V.}\ \bibnamefont
  {Lyubon'ko}}, \bibinfo {author} {\bibfnamefont {A.~Y.}\ \bibnamefont
  {Nechiporuk}}, \bibinfo {author} {\bibfnamefont {S.}~\bibnamefont
  {Ryabchenko}},  \emph {et~al.},\ }\href@noop {} {\bibfield  {journal}
  {\bibinfo  {journal} {Soviet Physics Journal}\ }\textbf {\bibinfo {volume}
  {32}},\ \bibinfo {pages} {276} (\bibinfo {year} {1989})}\BibitemShut
  {NoStop}%
\bibitem [{\citenamefont {Huebl}\ \emph {et~al.}(2013)\citenamefont {Huebl},
  \citenamefont {Zollitsch}, \citenamefont {Lotze}, \citenamefont {Hocke},
  \citenamefont {Greifenstein}, \citenamefont {Marx}, \citenamefont {Gross},\
  and\ \citenamefont {Goennenwein}}]{Huebl2013}%
  \BibitemOpen
  \bibfield  {author} {\bibinfo {author} {\bibfnamefont {H.}~\bibnamefont
  {Huebl}}, \bibinfo {author} {\bibfnamefont {C.~W.}\ \bibnamefont
  {Zollitsch}}, \bibinfo {author} {\bibfnamefont {J.}~\bibnamefont {Lotze}},
  \bibinfo {author} {\bibfnamefont {F.}~\bibnamefont {Hocke}}, \bibinfo
  {author} {\bibfnamefont {M.}~\bibnamefont {Greifenstein}}, \bibinfo {author}
  {\bibfnamefont {A.}~\bibnamefont {Marx}}, \bibinfo {author} {\bibfnamefont
  {R.}~\bibnamefont {Gross}}, \ and\ \bibinfo {author} {\bibfnamefont {S.~T.}\
  \bibnamefont {Goennenwein}},\ }\href@noop {} {\bibfield  {journal} {\bibinfo
  {journal} {Physical Review Letters}\ }\textbf {\bibinfo {volume} {111}},\
  \bibinfo {pages} {127003} (\bibinfo {year} {2013})}\BibitemShut {NoStop}%
\bibitem [{\citenamefont {Tabuchi}\ \emph {et~al.}(2014)\citenamefont
  {Tabuchi}, \citenamefont {Ishino}, \citenamefont {Ishikawa}, \citenamefont
  {Yamazaki}, \citenamefont {Usami},\ and\ \citenamefont
  {Nakamura}}]{Tabuchi2014}%
  \BibitemOpen
  \bibfield  {author} {\bibinfo {author} {\bibfnamefont {Y.}~\bibnamefont
  {Tabuchi}}, \bibinfo {author} {\bibfnamefont {S.}~\bibnamefont {Ishino}},
  \bibinfo {author} {\bibfnamefont {T.}~\bibnamefont {Ishikawa}}, \bibinfo
  {author} {\bibfnamefont {R.}~\bibnamefont {Yamazaki}}, \bibinfo {author}
  {\bibfnamefont {K.}~\bibnamefont {Usami}}, \ and\ \bibinfo {author}
  {\bibfnamefont {Y.}~\bibnamefont {Nakamura}},\ }\href {\doibase
  10.1103/PhysRevLett.113.083603} {\bibfield  {journal} {\bibinfo  {journal}
  {Phys. Rev. Lett.}\ }\textbf {\bibinfo {volume} {113}},\ \bibinfo {pages}
  {083603} (\bibinfo {year} {2014})}\BibitemShut {NoStop}%
\bibitem [{\citenamefont {Heyroth}\ \emph {et~al.}(2019)\citenamefont
  {Heyroth}, \citenamefont {Hauser}, \citenamefont {Trempler}, \citenamefont
  {Geyer}, \citenamefont {Syrowatka}, \citenamefont {Dreyer}, \citenamefont
  {Ebbinghaus}, \citenamefont {Woltersdorf},\ and\ \citenamefont
  {Schmidt}}]{Heyroth2019}%
  \BibitemOpen
  \bibfield  {author} {\bibinfo {author} {\bibfnamefont {F.}~\bibnamefont
  {Heyroth}}, \bibinfo {author} {\bibfnamefont {C.}~\bibnamefont {Hauser}},
  \bibinfo {author} {\bibfnamefont {P.}~\bibnamefont {Trempler}}, \bibinfo
  {author} {\bibfnamefont {P.}~\bibnamefont {Geyer}}, \bibinfo {author}
  {\bibfnamefont {F.}~\bibnamefont {Syrowatka}}, \bibinfo {author}
  {\bibfnamefont {R.}~\bibnamefont {Dreyer}}, \bibinfo {author} {\bibfnamefont
  {S.}~\bibnamefont {Ebbinghaus}}, \bibinfo {author} {\bibfnamefont
  {G.}~\bibnamefont {Woltersdorf}}, \ and\ \bibinfo {author} {\bibfnamefont
  {G.}~\bibnamefont {Schmidt}},\ }\href {\doibase
  10.1103/PhysRevApplied.12.054031} {\bibfield  {journal} {\bibinfo  {journal}
  {Phys. Rev. Applied}\ }\textbf {\bibinfo {volume} {12}},\ \bibinfo {pages}
  {054031} (\bibinfo {year} {2019})}\BibitemShut {NoStop}%
\bibitem [{\citenamefont {Mohammadi}\ \emph {et~al.}(2017)\citenamefont
  {Mohammadi}, \citenamefont {fazli}, \citenamefont {karevan},\ and\
  \citenamefont {Davoodi}}]{Mohammadi2017}%
  \BibitemOpen
  \bibfield  {author} {\bibinfo {author} {\bibfnamefont {M.}~\bibnamefont
  {Mohammadi}}, \bibinfo {author} {\bibfnamefont {H.}~\bibnamefont {fazli}},
  \bibinfo {author} {\bibfnamefont {M.}~\bibnamefont {karevan}}, \ and\
  \bibinfo {author} {\bibfnamefont {J.}~\bibnamefont {Davoodi}},\ }\href
  {\doibase https://doi.org/10.1016/j.eurpolymj.2017.03.056} {\bibfield
  {journal} {\bibinfo  {journal} {European Polymer Journal}\ }\textbf {\bibinfo
  {volume} {91}},\ \bibinfo {pages} {121 } (\bibinfo {year}
  {2017})}\BibitemShut {NoStop}%
\bibitem [{\citenamefont {Hansen}\ \emph {et~al.}(1974)\citenamefont {Hansen},
  \citenamefont {Röschmann},\ and\ \citenamefont
  {Tolksdorf}}]{HansenRoeschmannTolksdorf1974}%
  \BibitemOpen
  \bibfield  {author} {\bibinfo {author} {\bibfnamefont {P.}~\bibnamefont
  {Hansen}}, \bibinfo {author} {\bibfnamefont {P.}~\bibnamefont {Röschmann}},
  \ and\ \bibinfo {author} {\bibfnamefont {W.}~\bibnamefont {Tolksdorf}},\
  }\href {\doibase 10.1063/1.1663657} {\bibfield  {journal} {\bibinfo
  {journal} {Journal of Applied Physics}\ }\textbf {\bibinfo {volume} {45}},\
  \bibinfo {pages} {2728} (\bibinfo {year} {1974})},\ \Eprint
  {http://arxiv.org/abs/https://doi.org/10.1063/1.1663657}
  {https://doi.org/10.1063/1.1663657} \BibitemShut {NoStop}%
\bibitem [{\citenamefont {Tamaru}\ \emph {et~al.}(2002)\citenamefont {Tamaru},
  \citenamefont {Bain}, \citenamefont {Van~de Veerdonk}, \citenamefont
  {Crawford}, \citenamefont {Covington},\ and\ \citenamefont
  {Kryder}}]{Tamaru2002}%
  \BibitemOpen
  \bibfield  {author} {\bibinfo {author} {\bibfnamefont {S.}~\bibnamefont
  {Tamaru}}, \bibinfo {author} {\bibfnamefont {J.}~\bibnamefont {Bain}},
  \bibinfo {author} {\bibfnamefont {R.}~\bibnamefont {Van~de Veerdonk}},
  \bibinfo {author} {\bibfnamefont {T.}~\bibnamefont {Crawford}}, \bibinfo
  {author} {\bibfnamefont {M.}~\bibnamefont {Covington}}, \ and\ \bibinfo
  {author} {\bibfnamefont {M.}~\bibnamefont {Kryder}},\ }\href@noop {}
  {\bibfield  {journal} {\bibinfo  {journal} {Journal of Applied Physics}\
  }\textbf {\bibinfo {volume} {91}},\ \bibinfo {pages} {8034} (\bibinfo {year}
  {2002})}\BibitemShut {NoStop}%
\bibitem [{\citenamefont {Neudecker}\ \emph {et~al.}(2006)\citenamefont
  {Neudecker}, \citenamefont {Woltersdorf}, \citenamefont {Heinrich},
  \citenamefont {Okuno}, \citenamefont {Gubbiotti},\ and\ \citenamefont
  {Back}}]{Neudecker2006}%
  \BibitemOpen
  \bibfield  {author} {\bibinfo {author} {\bibfnamefont {I.}~\bibnamefont
  {Neudecker}}, \bibinfo {author} {\bibfnamefont {G.}~\bibnamefont
  {Woltersdorf}}, \bibinfo {author} {\bibfnamefont {B.}~\bibnamefont
  {Heinrich}}, \bibinfo {author} {\bibfnamefont {T.}~\bibnamefont {Okuno}},
  \bibinfo {author} {\bibfnamefont {G.}~\bibnamefont {Gubbiotti}}, \ and\
  \bibinfo {author} {\bibfnamefont {C.}~\bibnamefont {Back}},\ }\href@noop {}
  {\bibfield  {journal} {\bibinfo  {journal} {Journal of Magnetism and Magnetic
  Materials}\ }\textbf {\bibinfo {volume} {307}},\ \bibinfo {pages} {148}
  (\bibinfo {year} {2006})}\BibitemShut {NoStop}%
\bibitem [{\citenamefont {d'Allivy Kelly}\ \emph {et~al.}(2013)\citenamefont
  {d'Allivy Kelly}, \citenamefont {Anane}, \citenamefont {Bernard},
  \citenamefont {Ben~Youssef}, \citenamefont {Hahn}, \citenamefont
  {Molpeceres}, \citenamefont {Carr{\'e}t{\'e}ro}, \citenamefont {Jacquet},
  \citenamefont {Deranlot}, \citenamefont {Bortolotti} \emph
  {et~al.}}]{Kelly2013}%
  \BibitemOpen
  \bibfield  {author} {\bibinfo {author} {\bibfnamefont {O.}~\bibnamefont
  {d'Allivy Kelly}}, \bibinfo {author} {\bibfnamefont {A.}~\bibnamefont
  {Anane}}, \bibinfo {author} {\bibfnamefont {R.}~\bibnamefont {Bernard}},
  \bibinfo {author} {\bibfnamefont {J.}~\bibnamefont {Ben~Youssef}}, \bibinfo
  {author} {\bibfnamefont {C.}~\bibnamefont {Hahn}}, \bibinfo {author}
  {\bibfnamefont {A.~H.}\ \bibnamefont {Molpeceres}}, \bibinfo {author}
  {\bibfnamefont {C.}~\bibnamefont {Carr{\'e}t{\'e}ro}}, \bibinfo {author}
  {\bibfnamefont {E.}~\bibnamefont {Jacquet}}, \bibinfo {author} {\bibfnamefont
  {C.}~\bibnamefont {Deranlot}}, \bibinfo {author} {\bibfnamefont
  {P.}~\bibnamefont {Bortolotti}},  \emph {et~al.},\ }\href@noop {} {\bibfield
  {journal} {\bibinfo  {journal} {Applied Physics Letters}\ }\textbf {\bibinfo
  {volume} {103}},\ \bibinfo {pages} {082408} (\bibinfo {year}
  {2013})}\BibitemShut {NoStop}%
\bibitem [{\citenamefont {Ding}\ \emph {et~al.}(2020)\citenamefont {Ding},
  \citenamefont {Liu}, \citenamefont {Chang},\ and\ \citenamefont
  {Wu}}]{Ding2020}%
  \BibitemOpen
  \bibfield  {author} {\bibinfo {author} {\bibfnamefont {J.}~\bibnamefont
  {Ding}}, \bibinfo {author} {\bibfnamefont {T.}~\bibnamefont {Liu}}, \bibinfo
  {author} {\bibfnamefont {H.}~\bibnamefont {Chang}}, \ and\ \bibinfo {author}
  {\bibfnamefont {M.}~\bibnamefont {Wu}},\ }\href@noop {} {\bibfield  {journal}
  {\bibinfo  {journal} {IEEE Magnetics Letters}\ }\textbf {\bibinfo {volume}
  {11}},\ \bibinfo {pages} {1} (\bibinfo {year} {2020})}\BibitemShut {NoStop}%
\bibitem [{\citenamefont {Li}\ \emph {et~al.}(2016)\citenamefont {Li},
  \citenamefont {Xu}, \citenamefont {Aldosary}, \citenamefont {Tang},
  \citenamefont {Lin}, \citenamefont {Zhang}, \citenamefont {Lake},\ and\
  \citenamefont {Shi}}]{Li2016}%
  \BibitemOpen
  \bibfield  {author} {\bibinfo {author} {\bibfnamefont {J.}~\bibnamefont
  {Li}}, \bibinfo {author} {\bibfnamefont {Y.}~\bibnamefont {Xu}}, \bibinfo
  {author} {\bibfnamefont {M.}~\bibnamefont {Aldosary}}, \bibinfo {author}
  {\bibfnamefont {C.}~\bibnamefont {Tang}}, \bibinfo {author} {\bibfnamefont
  {Z.}~\bibnamefont {Lin}}, \bibinfo {author} {\bibfnamefont {S.}~\bibnamefont
  {Zhang}}, \bibinfo {author} {\bibfnamefont {R.}~\bibnamefont {Lake}}, \ and\
  \bibinfo {author} {\bibfnamefont {J.}~\bibnamefont {Shi}},\ }\href@noop {}
  {\bibfield  {journal} {\bibinfo  {journal} {Nature communications}\ }\textbf
  {\bibinfo {volume} {7}},\ \bibinfo {pages} {1} (\bibinfo {year}
  {2016})}\BibitemShut {NoStop}%
\bibitem [{\citenamefont {van Loo}\ \emph {et~al.}(2018)\citenamefont {van
  Loo}, \citenamefont {Morris},\ and\ \citenamefont {Karenowska}}]{Loo2018}%
  \BibitemOpen
  \bibfield  {author} {\bibinfo {author} {\bibfnamefont {A.~F.}\ \bibnamefont
  {van Loo}}, \bibinfo {author} {\bibfnamefont {R.~G.~E.}\ \bibnamefont
  {Morris}}, \ and\ \bibinfo {author} {\bibfnamefont {A.~D.}\ \bibnamefont
  {Karenowska}},\ }\href {\doibase 10.1103/PhysRevApplied.10.044070} {\bibfield
   {journal} {\bibinfo  {journal} {Phys. Rev. Applied}\ }\textbf {\bibinfo
  {volume} {10}},\ \bibinfo {pages} {044070} (\bibinfo {year}
  {2018})}\BibitemShut {NoStop}%
\bibitem [{\citenamefont {Mihalceanu}\ \emph {et~al.}(2018)\citenamefont
  {Mihalceanu}, \citenamefont {Vasyuchka}, \citenamefont {Bozhko},
  \citenamefont {Langner}, \citenamefont {Nechiporuk}, \citenamefont
  {Romanyuk}, \citenamefont {Hillebrands},\ and\ \citenamefont
  {Serga}}]{Mihalceanu2018}%
  \BibitemOpen
  \bibfield  {author} {\bibinfo {author} {\bibfnamefont {L.}~\bibnamefont
  {Mihalceanu}}, \bibinfo {author} {\bibfnamefont {V.~I.}\ \bibnamefont
  {Vasyuchka}}, \bibinfo {author} {\bibfnamefont {D.~A.}\ \bibnamefont
  {Bozhko}}, \bibinfo {author} {\bibfnamefont {T.}~\bibnamefont {Langner}},
  \bibinfo {author} {\bibfnamefont {A.~Y.}\ \bibnamefont {Nechiporuk}},
  \bibinfo {author} {\bibfnamefont {V.~F.}\ \bibnamefont {Romanyuk}}, \bibinfo
  {author} {\bibfnamefont {B.}~\bibnamefont {Hillebrands}}, \ and\ \bibinfo
  {author} {\bibfnamefont {A.~A.}\ \bibnamefont {Serga}},\ }\href {\doibase
  10.1103/PhysRevB.97.214405} {\bibfield  {journal} {\bibinfo  {journal} {Phys.
  Rev. B}\ }\textbf {\bibinfo {volume} {97}},\ \bibinfo {pages} {214405}
  (\bibinfo {year} {2018})}\BibitemShut {NoStop}%
\bibitem [{\citenamefont {Demokritov}\ \emph {et~al.}(2006)\citenamefont
  {Demokritov}, \citenamefont {Demidov}, \citenamefont {Dzyapko}, \citenamefont
  {Melkov}, \citenamefont {Serga}, \citenamefont {Hillebrands},\ and\
  \citenamefont {Slavin}}]{Demokritov2006}%
  \BibitemOpen
  \bibfield  {author} {\bibinfo {author} {\bibfnamefont {S.~O.}\ \bibnamefont
  {Demokritov}}, \bibinfo {author} {\bibfnamefont {V.~E.}\ \bibnamefont
  {Demidov}}, \bibinfo {author} {\bibfnamefont {O.}~\bibnamefont {Dzyapko}},
  \bibinfo {author} {\bibfnamefont {G.~A.}\ \bibnamefont {Melkov}}, \bibinfo
  {author} {\bibfnamefont {A.~A.}\ \bibnamefont {Serga}}, \bibinfo {author}
  {\bibfnamefont {B.}~\bibnamefont {Hillebrands}}, \ and\ \bibinfo {author}
  {\bibfnamefont {A.~N.}\ \bibnamefont {Slavin}},\ }\href@noop {} {\bibfield
  {journal} {\bibinfo  {journal} {Nature}\ }\textbf {\bibinfo {volume} {443}},\
  \bibinfo {pages} {430} (\bibinfo {year} {2006})}\BibitemShut {NoStop}%
\end{thebibliography}
\end{document}